\documentclass[reprint,superscriptaddress,amsmath,amssymb,aps,prl]{revtex4-2}
\usepackage{amsmath, amsfonts}
\usepackage{hyperref}
\usepackage{xr-hyper}
\usepackage{float}
\usepackage{graphicx}
\usepackage{dcolumn}
\usepackage{bm}
\usepackage{xcolor}
\usepackage{multirow}
\usepackage{atbegshi}
\usepackage{booktabs}
\usepackage[sort&compress]{cleveref}
\crefname{figure}{FIG.}{FIG.}
\Crefname{figure}{FIG.}{FIG.}
\crefname{table}{TABLE.}{TABLE.}
\Crefname{table}{TABLE.}{TABLE.}

\begin{document}
\title{\textbf{Imaging the N\'eel Vector in Two-Dimensional Antiferromagnets \\ using Antisymmetric Compton Scattering}}
\author{Wuxuan Li}
\thanks{These authors contributed equally to this work.}
\affiliation{Center for Quantum Matter, School of Physics, Zhejiang University, Hangzhou 310058, China.}
\author{Zhuocheng Lu}
\thanks{These authors contributed equally to this work.}
\affiliation{Center for Quantum Matter, School of Physics, Zhejiang University, Hangzhou 310058, China.}
\author{Jingshan Qi}
\email{qijingshan@email.tjut.edu.cn}
\affiliation{Tianjin Key Laboratory of Quantum Optics and Intelligent Photonics, School of Science, Tianjin University of Technology, Tianjin 300384, China}
\author{Hua Wang}
\email{daodaohw@zju.edu.cn}
\affiliation{Center for Quantum Matter, School of Physics, Zhejiang University, Hangzhou 310058, China.}
\author{Kai Chang}
\affiliation{Center for Quantum Matter, School of Physics, Zhejiang University, Hangzhou 310058, China.}
\date{\today}

\renewcommand{\abstractname}{}
\begin{abstract}
We demonstrate that antisymmetric Compton scattering can detect both the switching and the continuous rotation of the N\'eel vector in two-dimensional (2D) antiferromagnets. By probing magnetoelectric (ME) multipoles, which couple electric and magnetic dipoles, this approach overcomes the limitations of conventional techniques that rely on a finite net magnetization. Using a group-theoretical decomposition of the staggered moments in 2D MnPS$_3$ into irreducible representations, combined with first-principles calculations, we show that the antisymmetric Compton profile (ACP) is highly sensitive to the N\'eel vector orientation: it reverses sign under N\'eel vector reversal and exhibits distinct anisotropies under in-plane rotation. These results establish the ACP as a versatile probe of antiferromagnetic (AFM) order and magnetoelectric phenomena in van der Waals materials.
\end{abstract}
\maketitle

\textit{Introduction.}---The growing demand for miniaturized, lightweight, and high-performance spintronic devices that are resilient to electromagnetic interference is exposing the intrinsic limitations of conventional ferromagnet-based technologies. Two-dimensional (2D) antiferromagnetic (AFM) materials, in which neighboring magnetic moments align antiparallel to yield a vanishing net moment, have emerged as a compelling platform to address these challenges~\cite{neel1953some,nvemec2018antiferromagnetic,kim2019suppression,cheong2020seeing,gao2021layer}. The absence of stray-field crosstalk, ultrafast terahertz spin dynamics, intrinsic robustness against external magnetic perturbations, and remarkable tunability through van der Waals stacking, strain engineering, and ultrafast photomagnetic interactions~\cite{rondinelli2008carrier,hirohata2014future,spaldin2019advances,ahn20202d,thole2020concepts,shirazi2021magnetostrictive,xue2025reversing} together render them ideal candidates for next-generation, interference-resistant spintronics. Among the rapidly expanding family of 2D magnets, the transition-metal phosphorus trichalcogenides, $\mathrm{MnPX}_3$ ($\mathrm{X} = \mathrm{S,\,Se}$) are particularly attractive for magnetoelectric (ME) applications, owing to their facile exfoliation, intrinsic ME coupling, and the demonstrated strain and light controllability of the N\'eel vector~\cite{ressouche2010magnetoelectric,kim2019antiferromagnetic,long2020persistence,ni2021imaging}.

The very property that makes these compensated antiferromagnets technologically appealing---namely, their vanishing macroscopic magnetization---also renders their primary order parameter, the N\'eel vector $\mathbf{L}$, notoriously difficult to detect and image. Existing probes face an inherent trade-off between directness and accessibility. Neutron diffraction remains the gold standard for resolving magnetic unit cells and spin configurations, but it relies on large-scale facilities and bulk-like sample volumes that are incompatible with atomically thin specimens~\cite{urru2023neutron,taylor2011antiferromagnetic,muller2013grazing,jeevan2011muon}. Laboratory-based techniques such as second-harmonic generation (SHG)~\cite{fiebig2001second,reshak2017spin,sun2019giant}, layer and nonlinear Hall responses~\cite{tao2024layer,shao2020nonlinear}, magneto-transport~\cite{macdonald2011antiferromagnetic,du2023electrical,jungwirth2016antiferromagnetic}, and Bragg scattering~\cite{corcovilos2010detecting} efficiently report on symmetry breaking, yet they do not yield a direct reciprocal-space map of the underlying spin arrangement. A probe that combines momentum-space resolution with direct sensitivity to the N\'eel vector is therefore highly desirable.

In this Letter, we propose antisymmetric Compton scattering as such a probe of AFM order in 2D materials~\cite{bhowal2021revealing,cooper2004x,collins2016possibility,pratt2010compton}. Compton scattering, in which X-ray photons are inelastically scattered by electrons, has long served as a powerful momentum-space technique for interrogating the electronic structure of solids; more recently, it has been applied to ferro- and ferrimagnetic systems to extract spin-resolved electron momentum densities (EMD)~\cite{collins2016possibility}, hinting at its potential for probing N\'eel order~\cite{baltz2018antiferromagnetic,nvemec2018antiferromagnetic}. Building on this foundation, we show that the antisymmetric  Compton profile (ACP), which is the antisymmetric part of Compton profile, measurable via inelastic X-ray scattering of the EMD $\rho(\mathbf{p})$, acts as a selective and symmetry-protected fingerprint of the ME multipoles that characterize AFM order.

To establish this connection on a rigorous footing, we combine a group-theoretical classification of ME multipoles with first-principles calculations. We decompose AFM arrangements of magnetic moments into ME multipoles~\cite{spaldin2013monopole} associated with specific irreducible representations (irreps) of the crystallographic point group~\cite{watanabe2017magnetic,watanabe2018group}, and identify the symmetry channels in which the ACP is allowed. Taking monolayer $\mathrm{MnPS}_3$ as a representative prototype, we then show by explicit calculations that the ACP not only reverses sign under a $180^\circ$ flip of the N\'eel vector $\mathbf{L}$, but also responds sensitively to continuous in-plane rotations of $\mathbf{L}$, providing a one-to-one mapping between magnetic configuration and momentum-space response. Our results elevate the ACP from a largely overlooked component of the Compton profile to a versatile theoretical observable for imaging N\'eel order, with direct implications for studies of AFM phase transitions, magnetic domain dynamics, and, more broadly, emergent ME and altermagnetic phenomena in quantum materials.

\begin{figure}[t!]
\centering
\includegraphics[width=0.48\textwidth]{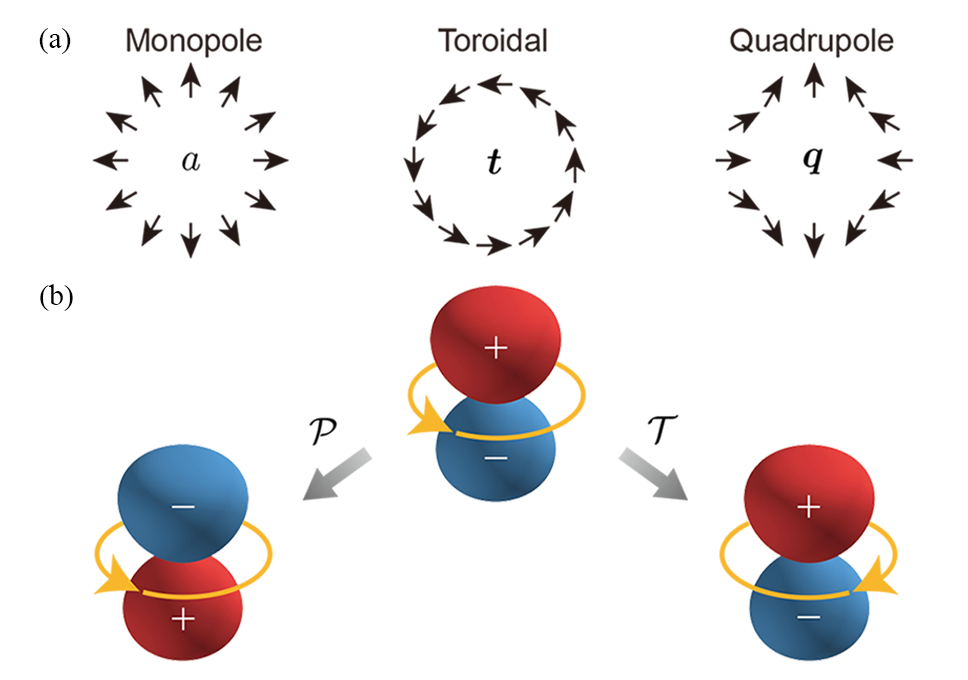}
\caption{(a) ME monopole, toroidal, and quadrupole moments constructed from spin moments (black arrows), corresponding to scalar, vector, and tensor forms, respectively. (b) Transformations of ME multipoles under $\mathcal{P}$ and $\mathcal{T}$: the polarization (blue/red spheres) is $\mathcal{T}$-even and $\mathcal{P}$-odd, whereas the magnetic moment (circular arrow) has the opposite parity.}
\label{f01}
\end{figure}

\textit{ME multipoles.}---ME multipoles arise from the multipole expansion of the interaction energy $\varepsilon_{\mathrm{int}}$ between a spatially varying external magnetic field $\boldsymbol{H}(\boldsymbol{r})$ and the magnetization density $\boldsymbol{\mu}(\boldsymbol{r})$~\cite{schaufelberger2023exploring,bhowal2021revealing,bhowal2022anti}:
\begin{equation}
\varepsilon_{\mathrm{int}} = -\mu_i H_i(0) - M_{ij}\,\partial_j H_i(0) - \mathcal{M}_{ijk} \partial_i\partial_j H_k(0).
\end{equation}
Here, $\mu_i = \int \mu_i(\boldsymbol{r})\,d^3r$ is the total magnetic dipole moment, $M_{ij} = \int \mu_i(\boldsymbol{r})\,r_j\,d^3r$ is the ME tensor, and $\mathcal{M}_{ijk} = \int \mu_k(\boldsymbol{r})\,r_i r_j\,d^3r$ denotes the octupole contribution. The first term couples the uniform field $H(0)$ to $\mu$, whereas the second couples the field gradient to $M_{ij}$. The ME tensor decomposes into three irreducible components~\cite{ederer2007towards,spaldin2013monopole,spaldin2021analogy,spaldin2008toroidal,thole2018magnetoelectric}: the monopole $a = \tfrac{1}{3} M_{kk}$, the toroidal moment $t_k = \tfrac{1}{2}\epsilon_{kij}(M_{ij}-M_{ji})$, and the traceless quadrupole $q_{ij} = \tfrac{1}{2}(M_{ij}+M_{ji}) - \tfrac{1}{3}\delta_{ij} M_{kk}$. Their schematic configurations are depicted in Fig.~\ref{f01}(a). Any magnetic alignment can be expanded as a linear combination of these ME multipoles, providing a unified framework to describe AFM order~\cite{ederer2007towards,spaldin2013monopole}. In contrast to magnetic dipoles, which are odd under time reversal ($\mathcal{T}$) but even under spatial inversion ($\mathcal{P}$), ME multipoles are odd under both $\mathcal{T}$ and $\mathcal{P}$. The N\'eel vector $\mathbf{L}$, defined as the difference of sublattice magnetizations, transforms identically under $\mathcal{P}$ and $\mathcal{T}$ [Fig.~\ref{f01}(b)]. ME multipoles therefore provide a natural framework for the symmetry classification and detection of $\mathbf{L}$.

\textit{Group-theoretical classification.}---To establish a rigorous link between the ACP and magnetic order, we classify the ME multipoles using the irreps of the crystal point group~\cite{watanabe2018group,watanabe2017magnetic}. ME multipoles, built from the product of displacement $r_i$ and magnetization $\mu_j$, transform under specific irreps that dictate their selection rules. For the $D_{3d}$ point group relevant to MnPS$_3$~\cite{malard2009group}, ME multipoles must lie in the parity-odd irreps ($A_{1u}$, $A_{2u}$, $E_{u}$). For instance, the magnetic quadrupole component $q_{xy} = x\mu_y + y\mu_x$ arises from the product of displacement ($E_u$) and magnetization ($E_g$); the decomposition $E_u \otimes E_g = A_{1u} \oplus A_{2u} \oplus E_{u}$ confirms that these multipoles are symmetry-allowed and transform consistently within the group. A finer classification follows from compatibility relations with the parent $D_{6h}$ group~\cite{inui2012group,perez2015symmetry,bhowal2021revealing}. Under $D_{6h} \downarrow D_{3d}$, the ME monopole $a$ and the quadrupole $q_{z^2}$ transform as $A_{1u}$, the toroidal component $t_z$ as $A_{2u}$, and the remaining components as $E_{u}$ (Table~\ref{tab:D_3d_pg}). Because the ME multipoles share the same irreps as the N\'eel vector, they serve as direct, symmetry-protected proxies for the magnetic order. Characterizing them through the ACP therefore enables an unambiguous determination of both the magnitude and orientation of $\mathbf{L}$.

\begin{figure}[t]
\centering
\includegraphics[width=0.5\textwidth]{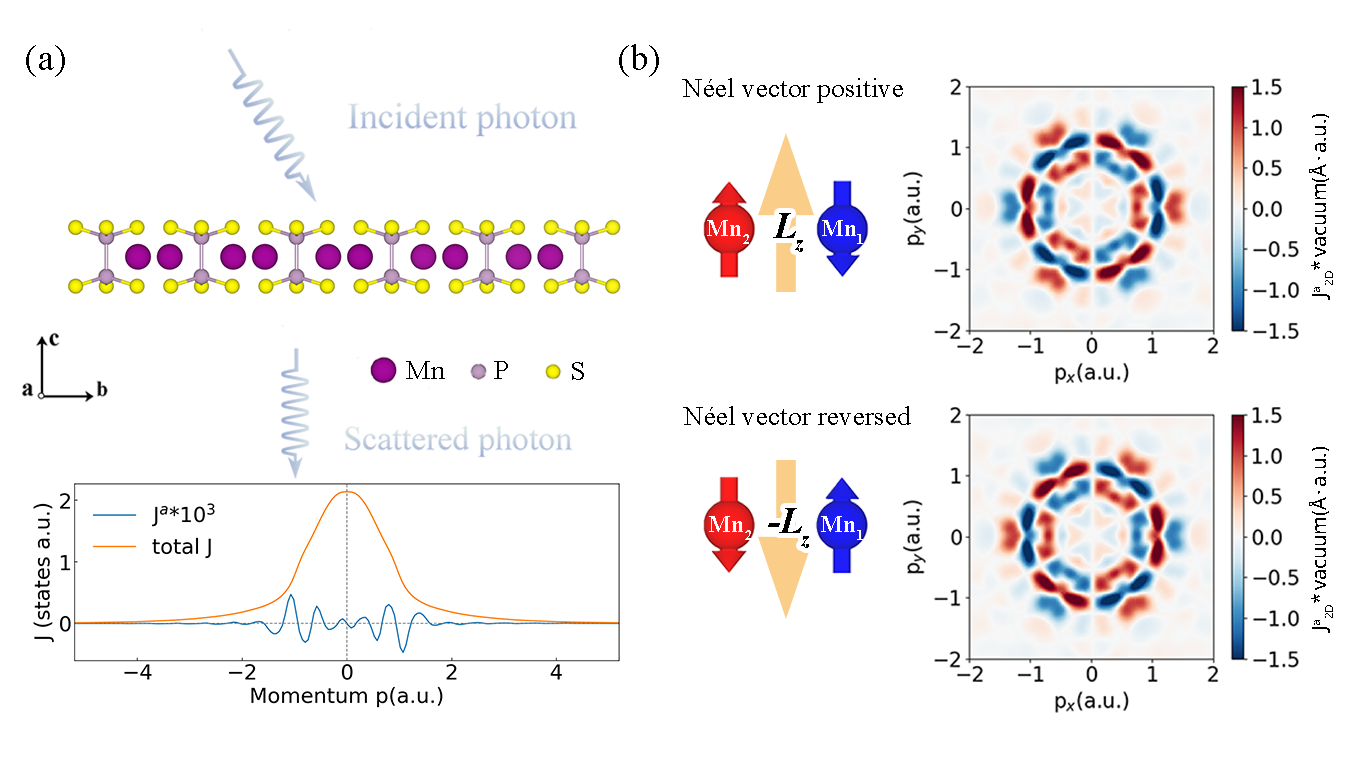}
\caption{(a) Atomic structure of monolayer MnPS$_3$. The total Compton profile (orange) and its antisymmetric component (blue) are compared. (b) Reversal of the N\'eel vector induces a sign change in the ACP intensity, evident from the color contrast. Two AFM configurations of MnPS$_3$ are shown, with N\'eel vectors pointing upward and downward, respectively. The ACP is magnified by a factor of $10^3$ in all panels.}
\label{f02}
\end{figure}

\textit{Antisymmetric Compton profile.}---The Compton profile $J(p_z)$ is extracted from the double-differential cross section of inelastically scattered photons and provides a momentum-space map of the electronic structure [Fig.~\ref{f02}(a)]~\cite{collins2016possibility,cooper1985compton}. Although $J(p_z)$ is generally symmetric, here we focus on its antisymmetric component, $J^a(p_z) = \tfrac{1}{2}[J(p_z) - J(-p_z)]$. Whenever spatial inversion ($\mathcal{P}$) or time reversal ($\mathcal{T}$) is individually preserved, the EMD obeys $\rho(\vec{p}) = \rho(-\vec{p})$, so that $J(p_z) = J(-p_z)$ and $J^a(p_z)$ vanishes. Simultaneous breaking of $\mathcal{P}$ and $\mathcal{T}$ while preserving their product $\mathcal{PT}$---the defining symmetry of many antiferromagnets---lifts this constraint and permits a finite $J^a(p_z)$. Crucially, $J^a(p_z)$ transforms under the same irrep as the ME multipoles and the N\'eel vector $\mathbf{L} = \mathbf{M}_1 - \mathbf{M}_2$~\cite{lin2020dirac,tao2024layer}. The ACP therefore serves as a direct, symmetry-protected fingerprint of AFM order, tracking the orientation of $\mathbf{L}$ through its momentum-space structure and signaling its reversal through a global sign change.

\begin{table}[b!]
\centering
\renewcommand{\arraystretch}{1.5}
\caption{Basis functions of the ME multipoles for the $D_{3d}$ point group: ME monopole $(a)$, toroidal moment $(\vec{t})$, and quadrupole moment $(q_{ij})$.}
\begin{tabular}{|c|c|c|c|}
\hline
\textbf{IRs} & ME multipoles & real-space basis & $k$-space basis \\ \hline
\multirow{2}{*}{$A_{1u}$}
& $a$        & $xm_x + ym_y + zm_z$ & \multirow{2}{*}{} \\
& $q_{z^2}$  & $2zm_z - xm_x - ym_y$ & \\ \hline
$A_{2u}$ & $t_z$ & $xm_y - ym_x$ & $k_z$ \\ \hline
\multirow{4}{*}{$E_u$}
& $\{t_x, t_y\}$       & $ym_z - zm_y,\ zm_x - xm_z$ & $\{k_x, k_y\}$ \\
& $\{q_{yz}, q_{xz}\}$ & $ym_z + zm_y,\ xm_z + zm_x$ & \\
& \multirow{2}{*}{$\{q_{xy}, q_{x^2-y^2}\}$} & \multirow{2}{*}{$xm_y + ym_x,\ xm_x - ym_y$} & $\{k_z(k_x^2-k_y^2),$ \\
&  &  & $k_x k_y k_z\}$ \\ \hline
\end{tabular}
\label{tab:D_3d_pg}
\end{table}

ACP measurements have already been used to reveal toroidal order and ME quadrupoles in bulk LiNiPO$_4$ and Mn$_2$Au~\cite{bhowal2021revealing}. In those high-symmetry orthorhombic ($D_{2h}$) and tetragonal ($D_{4h}$) systems, however, toroidal and quadrupolar contributions are intertwined. To disentangle the individual multipoles, we focus on the 2D semiconductor MnPS$_3$ ($D_{3d}$ point group), in which Mn$^{2+}$ ions form a honeycomb lattice with AFM nearest-neighbor coupling~\cite{ressouche2010magnetoelectric}. The lower symmetry of MnPS$_3$ permits a cleaner separation of the multipole contributions. To go beyond one-dimensional projections, we further consider the once-projected EMD $J_{2\mathrm{D}}(p_x, p_y) = \int \rho(\mathbf{p})\,dp_z$, often called the 2D-ACP, which provides a complete map of the electron distribution perpendicular to the scattering vector and is experimentally accessible via 2D angular correlation of annihilation radiation (2D-ACAR)~\cite{dugdale2014probing,mills1995positron}.

\textit{MnPS$_3$ monolayer.}---We now use the MnPS$_3$ monolayer with several distinct AFM configurations to illustrate the sensitivity of the ACP to the N\'eel vector orientation. We first consider an out-of-plane N\'eel vector. Symmetry analysis shows that MnPS$_3$ retains the $D_{3d}$ point group, with the magnetic order transforming as the $A_{2u}$ irrep. From Table~\ref{tab:D_3d_pg}, $A_{2u}$ supports a nonzero ME toroidal moment $t_z$ whose momentum-space basis is $k_z$. Because the ACP and the ME multipoles share the same symmetry, the basis function $t_z$ also implies an antisymmetric ACP between $(0,0,k_z)$ and $(0,0,-k_z)$, which are related by individual $\mathcal{P}$ or $\mathcal{T}$.

\begin{figure}[t]
\centering
\includegraphics[width=0.50\textwidth]{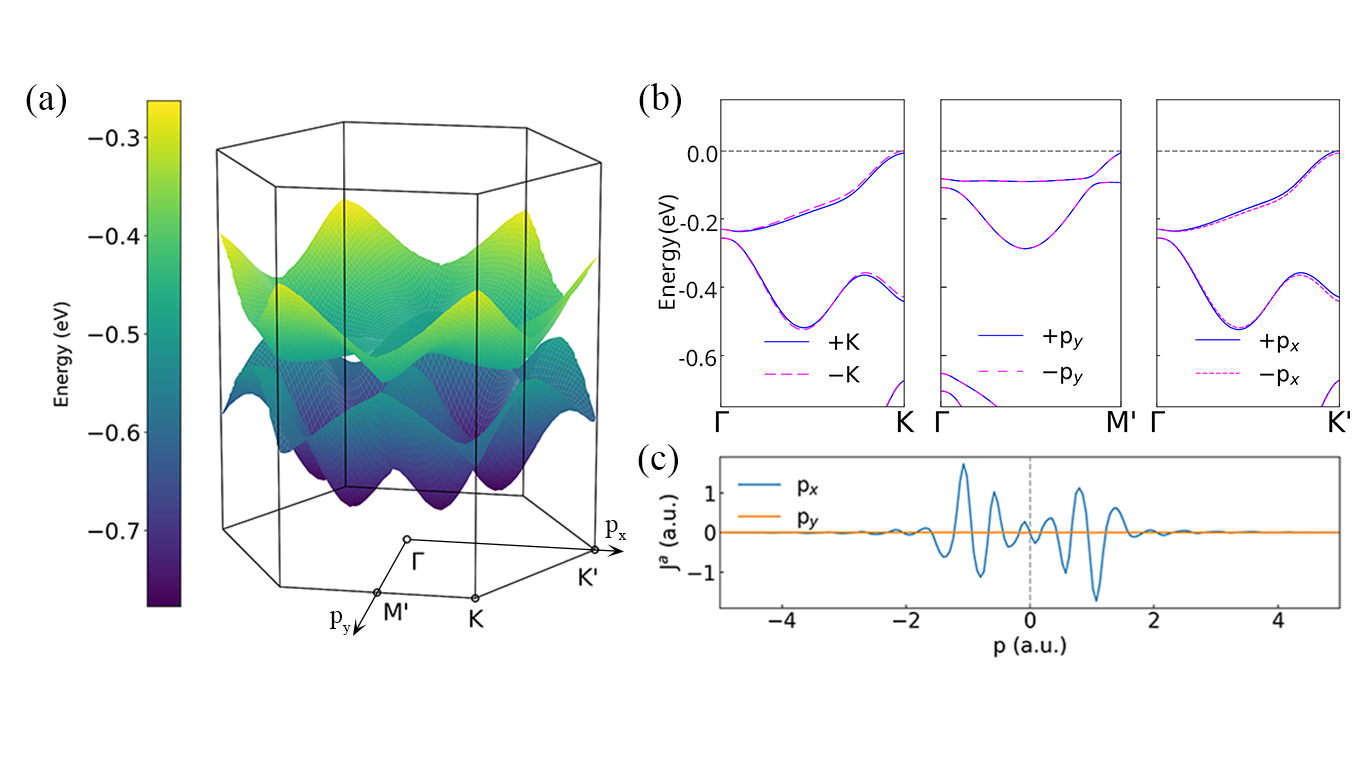}
\caption{(a) Three-dimensional conduction and valence bands of MnPS$_3$ within the first Brillouin zone (BZ). (b) Band structure along $\Gamma \!\to\! \mathrm{K'}$ ($\pm p_x$), $\Gamma \!\to\! \mathrm{M'}$ ($\pm p_y$), and $\Gamma \!\to\! \mathrm{K}$. (c) One-dimensional ACP along $\pm p_x$ (blue) and $\pm p_y$ (orange).}
\label{f03}
\end{figure}

Having established the symmetry properties of the ACP, we now examine their manifestation in first-principles calculations of MnPS$_3$ in the $p_x$--$p_y$ plane. As shown in the upper panel of Fig.~\ref{f02}(b), the calculated 2D-ACP for the out-of-plane AFM configuration exhibits a characteristic spiderweb-like pattern. The pattern displays threefold rotational symmetry, consistent with the $C_3$ and two $S_6$ axes along $z$. Three antisymmetric axes lie at $30^\circ$, $90^\circ$ ($p_y$), and $150^\circ$, in agreement with the antisymmetric $C_2$ axes perpendicular to the principal axis, while the symmetric axes at $0^\circ$ ($p_x$), $60^\circ$, and $120^\circ$ correspond to the $\sigma_d$ mirrors. This symmetry agreement confirms that the ACP not only reflects the underlying crystal symmetry but also serves as a sensitive probe of the ME multipole order. To assess sensitivity to N\'eel vector reversal, we flip all magnetic moments by $180^\circ$. The magnetic symmetry remains the $A_{2u}$ irrep of $D_{3d}$, and the resulting ACP, shown in the lower panel of Fig.~\ref{f02}(b), preserves the overall pattern symmetry but undergoes a global sign reversal (color inversion). The ACP thus provides a direct imaging tool for the N\'eel vector.

The momentum-space asymmetry encoded in the ACP can be related directly to the underlying electronic band structure: both are asymmetric under individual $\mathcal{P}$ or $\mathcal{T}$, while symmetric under combined $\mathcal{PT}$~\cite{bhowal2021revealing,bhowal2022hidden}. Each antisymmetric direction in the ACP corresponds to a momentum-space asymmetry in the bands. Figure~\ref{f03}(a) shows the band structure within the first BZ centered at $\Gamma$, where reversed high-symmetry paths are related by individual $\mathcal{P}$ or $\mathcal{T}$. Figure~\ref{f03}(b) presents three representative paths, $\Gamma \!\to\! \pm \mathrm{K}$, $\Gamma \!\to\! \pm p_y$, and $\Gamma \!\to\! \pm p_x$, corresponding respectively to the $150^\circ$-antisymmetric, $p_y$-symmetric, and $p_x$-antisymmetric axes of the ACP. The bands along these paths inherit the same symmetry properties as the corresponding ACP directions. Figure~\ref{f03}(c) compares the ACP magnitudes along $p_x$ and $p_y$, demonstrating that the ACP offers enhanced sensitivity to the direction of momentum-space asymmetry.

\begin{figure}[t]
\centering
\includegraphics[width=0.5\textwidth]{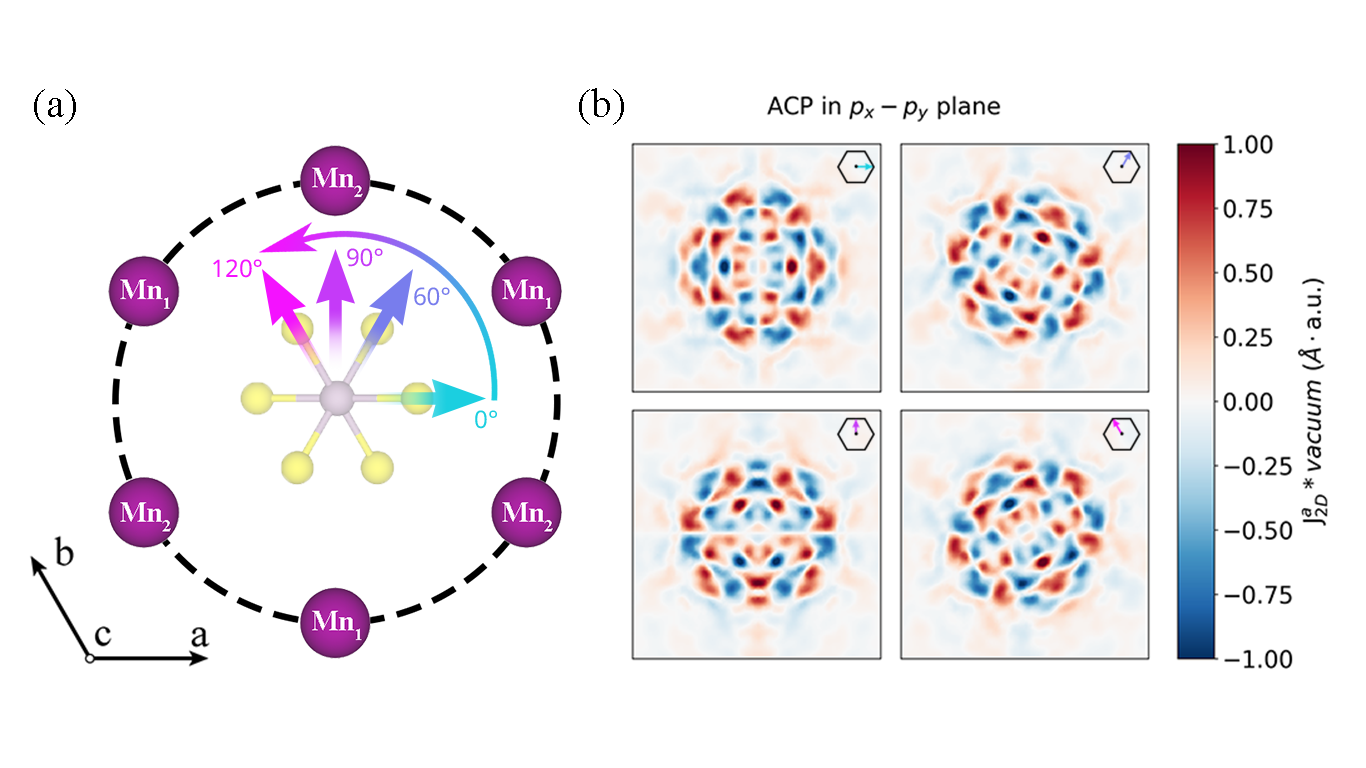}
\caption{2D-ACP under rotation of the N\'eel vector. Magenta arrows in (a) indicate the N\'eel vector orientation. The N\'eel vector is rotated from the lattice vector $\vec{a}$ to $\vec{b}$ by $120^\circ$. (b) Representative ACP patterns at $0^\circ$, $60^\circ$, $90^\circ$, and $120^\circ$ illustrate the rotation of the Compton profile.}
\label{f04}
\end{figure}

The correspondence between the ACP and the magnetic symmetry extends beyond simple N\'eel vector reversal: the modified symmetry of in-plane AFM configurations reshapes the momentum-space response and yields distinct ACP patterns. For an in-plane N\'eel vector, the symmetry reduces to the $C_{2h}$ point group~\cite{bhowal2021revealing}. Applying compatibility relations between $D_{4h}$ and its subgroup $C_{2h}$, the $E_u$ irrep of $D_{4h}$ reduces to $B_u$ of $C_{2h}$, while $A_{1u}$, $A_{2u}$, $B_{1u}$, and $B_{2u}$ all reduce to $A_u$. The resulting basis functions and corresponding ME multipoles in $C_{2h}$ are listed in Table~\ref{tab:C_2h_pg}. Symmetry analysis shows that an in-plane AFM order belongs to the $B_u$ irrep, which carries toroidal components $t_x, t_y$ and quadrupole components $q_{yz}, q_{xz}$ with $k$-space bases along $k_x$ and $k_y$. A nonzero ACP is therefore expected along both $k_x$ and $k_y$. Continuous rotation of the N\'eel vector within the $p_x$--$p_y$ plane provides a direct probe of the orientation dependence of the ACP. As illustrated in Fig.~\ref{f04}(a), the N\'eel vector rotates from $\vec{a}$ ([100]) to $\vec{b}$ ([010]) by $120^\circ$, with representative orientations at $0^\circ$ (along $\vec{a}$), $60^\circ$, $90^\circ$, and $120^\circ$ (along $\vec{b}$).

Correspondingly, the first-principles results in Fig.~\ref{f04}(b) display symmetry-dependent ACP patterns for distinct in-plane AFM configurations. The pattern symmetries match the assignments of the $B_u$ irrep of $C_{2h}$ given in the character table (see Supplemental Material Sec.~A and Table~S1). Colored dashed arrows mark the antisymmetric axes of the corresponding ACP, while black dotted lines denote symmetric axes. Comparison with the colored arrows in Fig.~\ref{f04}(a) shows that arrows of the same color have identical orientations, demonstrating that the antisymmetric axes of the 2D-ACP provide a direct visualization of the N\'eel vector orientation. Comparison with the band structure (Supplemental Material Sec.~C and Fig.~S2) further confirms that the directions of band asymmetry coincide with those along which the ACP is nonzero. The ACP thus not only detects N\'eel vector reversal but also responds sensitively to incremental rotations, providing a powerful probe of AFM states.

\begin{table}[t!]
\centering
\renewcommand{\arraystretch}{1.5}
\caption{Basis functions of the ME multipoles for the $C_{2h}$ point group: ME monopole $(a)$, toroidal moment $(\vec{t})$, and quadrupole moment $(q_{ij})$.}
\begin{tabular}{|c|c|c|c|}
\hline
\textbf{IRs} & ME multipoles & real-space basis & $k$-space basis \\ \hline
\multirow{4}{*}{$A_u$}
& $a$            & $xm_x + ym_y + zm_z$           & $k_x k_y k_z$ \\
& $t_z, q_{xy}$  & $xm_y - ym_x,\ xm_y + ym_x$    & $k_z$ \\
& $q_{x^2-y^2}$  & $xm_x - ym_y$                  & \\
& $q_{z^2}$      & $2zm_z - xm_x - ym_y$          & \\ \hline
\multirow{2}{*}{$B_u$}
& $t_x, q_{yz}$  & $ym_z - zm_y,\ ym_z + zm_y$    & $k_x$ \\
& $t_y, q_{xz}$  & $zm_x - xm_z,\ zm_x + xm_z$    & $k_y$ \\ \hline
\end{tabular}
\label{tab:C_2h_pg}
\end{table}

\textit{Conclusion.}---We have shown that the antisymmetric Compton profile arises from the symmetry-governed momentum-space asymmetry induced by the $\mathcal{P}$- and $\mathcal{T}$-breaking ME multipoles associated with the N\'eel vector in 2D antiferromagnets. Our \textit{ab initio} results for MnPS$_3$ confirm that the ACP is an ideal observable for tracking both the switching and the continuous rotation of the N\'eel vector. While experimentally resolving such a small signal requires high-brilliance X-ray sources combined with advanced data-processing techniques such as Compton tomography, our work provides the theoretical foundation for laboratory-based probes of magnetic order in quantum materials.

Building on the compatibility relations, we have further extended the symmetry analysis of the ME multipoles to all crystallographic point groups, with detailed tables provided in the Supplemental Material (Sec.~D, Tables~S2--S32). This systematic catalog offers a practical guide for identifying symmetry-allowed ME multipoles across diverse material classes. Looking forward, particular attention should be paid to higher-rank multipoles, such as magnetic octupoles $\mathcal{M}_{ijk}$ and hexadecapoles $\mathcal{M}_{ijkl}$~\cite{vsmejkal2022beyond,vsmejkal2022emerging,bhowal2024ferroically,bandyopadhyay2025designing}, and to their possible roles in altermagnetism, where suitable experimental probes are still scarce. By establishing the connection between the ACP and ME multipoles, our work delivers both a developed methodology to probe N\'eel order and a foundation for exploring emergent altermagnetic phases and higher-rank multipoles, opening new avenues for uncovering complex magnetoelectric phenomena in quantum materials.

\indent\textit{Acknowledgments.}---H.W. acknowledges support from the NSFC under Grants No.~12474240, No.~12522411, and No.~12304049, as well as support from the Fundamental Research Funds for the Central Universities. J.Q. acknowledges support from the NSFC under Grant No.~12574273. K.C. acknowledges support from the NSFC under Grants No.~92265203 and No.~12488101, and the Innovation Program for Quantum Science and Technology under Grant No.~2024ZD0300104.

\bibliography{Ref}

@article{neel1953some,
  title={Some new results on antiferromagnetism and ferromagnetism},
  author={N{\'e}el, Louis},
  journal={Reviews of Modern Physics},
  volume={25},
  number={1},
  pages={58},
  year={1953},
  publisher={APS}
}

@article{macdonald2011antiferromagnetic,
  title={Antiferromagnetic metal spintronics},
  author={MacDonald, AH and Tsoi, M},
  journal={Philosophical Transactions of the Royal Society A: Mathematical, Physical and Engineering Sciences},
  volume={369},
  number={1948},
  pages={3098--3114},
  year={2011},
  publisher={The Royal Society}
}

@article{baltz2018antiferromagnetic,
  title={Antiferromagnetic spintronics},
  author={Baltz, Vincent and Manchon, Aurelien and Tsoi, M and Moriyama, Takahiro and Ono, T and Tserkovnyak, Y},
  journal={Reviews of Modern Physics},
  volume={90},
  number={1},
  pages={015005},
  year={2018},
  publisher={APS}
}

@article{urru2023neutron,
  title={Neutron scattering from local magnetoelectric multipoles: A combined theoretical, computational, and experimental perspective},
  author={Urru, Andrea and Soh, Jian Rui and Qureshi, Navid and Stunault, Anne and Roessli, Bertrand and R{\o}nnow, Henrik M and Spaldin, Nicola A},
  journal={Physical Review Research},
  volume={5},
  number={3},
  pages={033147},
  year={2023},
  publisher={APS}
}

@article{ederer2007towards,
  title={Towards a microscopic theory of toroidal moments in bulk periodic crystals},
  author={Ederer, Claude and Spaldin, Nicola A},
  journal={Physical Review B},
  volume={76},
  number={21},
  pages={214404},
  year={2007},
  publisher={APS}
}

@book{cooper2004x,
  title={X-ray Compton scattering},
  author={Cooper, Malcolm and Mijnarends, Peter and Shiotani, Nobuhiro and Sakai, Nobuhiko and Bansil, Arun},
  volume={5},
  year={2004},
  publisher={OUP Oxford}
}

@article{pratt2010compton,
  title={Compton scattering revisited},
  author={Pratt, RH and LaJohn, LA and Florescu, V and Suri{\'c}, Tihomir and Chatterjee, Barun Kumar and Roy, SC},
  journal={Radiation Physics and Chemistry},
  volume={79},
  number={2},
  pages={124--131},
  year={2010},
  publisher={Elsevier}
}

@article{watanabe2017magnetic,
  title={Magnetic hexadecapole order and magnetopiezoelectric metal state in Ba 1- x K x Mn 2 As 2},
  author={Watanabe, Hikaru and Yanase, Youichi},
  journal={Physical Review B},
  volume={96},
  number={6},
  pages={064432},
  year={2017},
  publisher={APS}
}

@article{watanabe2018group,
  title={Group-theoretical classification of multipole order: Emergent responses and candidate materials},
  author={Watanabe, Hikaru and Yanase, Youichi},
  journal={Physical Review B},
  volume={98},
  number={24},
  pages={245129},
  year={2018},
  publisher={APS}
}

@article{schaufelberger2023exploring,
  title={Exploring energy landscapes of charge multipoles using constrained density functional theory},
  author={Schaufelberger, Luca and Merkel, Maximilian E and Tehrani, Aria Mansouri and Spaldin, Nicola A and Ederer, Claude},
  journal={Physical Review Research},
  volume={5},
  number={3},
  pages={033172},
  year={2023},
  publisher={APS}
}

@article{fiebig2001second,
  title={Second harmonic generation in the centrosymmetric antiferromagnet NiO},
  author={Fiebig, Manfred and Fr{\"o}hlich, D and Lottermoser, Th and Pavlov, VV and Pisarev, RV and Weber, HJ},
  journal={Physical Review Letters},
  volume={87},
  number={13},
  pages={137202},
  year={2001},
  publisher={APS}
}

@article{sun2019giant,
  title={Giant nonreciprocal second-harmonic generation from antiferromagnetic bilayer CrI3},
  author={Sun, Zeyuan and Yi, Yangfan and Song, Tiancheng and Clark, Genevieve and Huang, Bevin and Shan, Yuwei and Wu, Shuang and Huang, Di and Gao, Chunlei and Chen, Zhanghai and others},
  journal={Nature},
  volume={572},
  number={7770},
  pages={497--501},
  year={2019},
  publisher={Nature Publishing Group UK London}
}

@article{spaldin2013monopole,
  title={Monopole-based formalism for the diagonal magnetoelectric response},
  author={Spaldin, Nicola A and Fechner, Michael and Bousquet, Eric and Balatsky, Alexander and Nordstr{\"o}m, Lars},
  journal={Physical Review B},
  volume={88},
  number={9},
  pages={094429},
  year={2013},
  publisher={APS}
}

@article{nvemec2018antiferromagnetic,
  title={Antiferromagnetic opto-spintronics},
  author={N{\v{e}}mec, Petr and Fiebig, Manfred and Kampfrath, Tobias and Kimel, Alexey V},
  journal={Nature Physics},
  volume={14},
  number={3},
  pages={229--241},
  year={2018},
  publisher={Nature Publishing Group UK London}
}

@article{bhowal2021revealing,
  title={Revealing hidden magnetoelectric multipoles using Compton scattering},
  author={Bhowal, Sayantika and Spaldin, Nicola A},
  journal={Physical Review Research},
  volume={3},
  number={3},
  pages={033185},
  year={2021},
  publisher={APS}
}

@article{collins2016possibility,
  title={On the possibility of using X-ray Compton scattering to study magnetoelectrical properties of crystals},
  author={Collins, SP and Laundy, D and Connolley, T and van der Laan, G and Fabrizi, F and Janssen, O and Cooper, MJ and Ebert, H and Mankovsky, S},
  journal={Foundations of Crystallography},
  volume={72},
  number={2},
  pages={197--205},
  year={2016},
  publisher={International Union of Crystallography}
}

@article{corcovilos2010detecting,
  title={Detecting antiferromagnetism of atoms in an optical lattice via optical Bragg scattering},
  author={Corcovilos, TA and Baur, SK and Hitchcock, JM and Mueller, EJ and Hulet, RG},
  journal={Physical Review A},
  volume={81},
  number={1},
  pages={013415},
  year={2010},
  publisher={APS}
}

@article{ni2021imaging,
  title={Imaging the N{\'e}el vector switching in the monolayer antiferromagnet MnPSe3 with strain-controlled Ising order},
  author={Ni, Zhuoliang and Haglund, AV and Wang, H ea and Xu, B and Bernhard, C and Mandrus, DG and Qian, X and Mele, EJ and Kane, CL and Wu, Liang},
  journal={Nature Nanotechnology},
  volume={16},
  number={7},
  pages={782--787},
  year={2021},
  publisher={Nature Publishing Group UK London}
}

@book{shirazi2021magnetostrictive,
  title={Magnetostrictive Ferri \& Antiferromagnetic Thin Films for Multiferroic Applications},
  author={Shirazi, Paymon},
  year={2021},
  publisher={University of California, Los Angeles}
}

@article{ressouche2010magnetoelectric,
  title={Magnetoelectric MnPS 3 as a candidate for ferrotoroidicity},
  author={Ressouche, Eric and Loire, Mickael and Simonet, Virginie and Ballou, Rafik and Stunault, Anne and Wildes, Andrew},
  journal={Physical Review B},
  volume={82},
  number={10},
  pages={100408},
  year={2010},
  publisher={APS}
}

@article{bhowal2022anti,
  title={Anti-symmetric Compton scattering in LiNiPO 4: Towards a direct probe of the magneto-electric multipole moment},
  author={Bhowal, Sayantika and O'Neill, Daniel and Fechner, Michael and Spaldin, Nicola A and Staub, Urs and Duffy, Jon and Collins, Stephen P},
  journal={Open Research Europe},
  volume={1},
  pages={132},
  year={2022}
}

@article{spaldin2021analogy,
  title={Analogy between the magnetic dipole moment at the surface of a magnetoelectric and the electric charge at the surface of a ferroelectric},
  author={Spaldin, Nicola A},
  journal={Journal of Experimental and Theoretical physics},
  volume={132},
  number={4},
  pages={493--505},
  year={2021},
  publisher={Springer}
}

@book{inui2012group,
  title={Group theory and its applications in physics},
  author={Inui, Teturo and Tanabe, Yukito and Onodera, Yositaka},
  year={2012},
  publisher={Springer Science \& Business Media}
}

@article{perez2015symmetry,
  title={Symmetry-based computational tools for magnetic crystallography},
  author={Perez-Mato, JM and Gallego, SV and Tasci, ES and Elcoro, LU{\.I}S and de la Flor, Gemma and Aroyo, MI},
  journal={Annual Review of Materials Research},
  volume={45},
  number={1},
  pages={217--248},
  year={2015},
  publisher={Annual Reviews}
}

@article{cheong2020seeing,
  title={Seeing is believing: visualization of antiferromagnetic domains},
  author={Cheong, Sang Wook and Fiebig, Manfred and Wu, Weida and Chapon, Laurent and Kiryukhin, Valery},
  journal={npj Quantum Materials},
  volume={5},
  number={1},
  pages={3},
  year={2020},
  publisher={Nature Publishing Group UK London}
}

@article{kim2019suppression,
  title={Suppression of magnetic ordering in XXZ-type antiferromagnetic monolayer NiPS3},
  author={Kim, Kangwon and Lim, Soo Yeon and Lee, Jae-Ung and Lee, Sungmin and Kim, Tae Yun and Park, Kisoo and Jeon, Gun Sang and Park, Cheol Hwan and Park, Je-Geun and Cheong, Hyeonsik},
  journal={Nature Communications},
  volume={10},
  number={1},
  pages={345},
  year={2019},
  publisher={Nature Publishing Group UK London}
}

@article{gao2021layer,
  title={Layer Hall effect in a 2D topological axion antiferromagnet},
  author={Gao, Anyuan and Liu, YuFei and Hu, Chaowei and Qiu, JianXiang and Tzschaschel, Christian and Ghosh, Barun and Ho, ShengChin and B{\'e}rub{\'e}, Damien and Chen, Rui and Sun, Haipeng and others},
  journal={Nature},
  volume={595},
  number={7868},
  pages={521--525},
  year={2021},
  publisher={Nature Publishing Group UK London}
}

@article{hirohata2014future,
  title={Future perspectives for spintronic devices},
  author={Hirohata, Atsufumi and Takanashi, Koki},
  journal={Journal of Physics D: Applied Physics},
  volume={47},
  number={19},
  pages={193001},
  year={2014},
  publisher={IOP Publishing}
}

@article{rondinelli2008carrier,
  title={Carrier-mediated magnetoelectricity in complex oxide heterostructures},
  author={Rondinelli, James M and Stengel, Massimiliano and Spaldin, Nicola A},
  journal={Nature Nanotechnology},
  volume={3},
  number={1},
  pages={46--50},
  year={2008},
  publisher={Nature Publishing Group UK London}
}

@article{thole2020concepts,
  title={Concepts from the linear magnetoelectric effect that might be useful for antiferromagnetic spintronics},
  author={Th{\"o}le, Florian and Keliri, Andriani and Spaldin, Nicola A},
  journal={Journal of Applied Physics},
  volume={127},
  number={21},
  year={2020},
  publisher={AIP Publishing}
}

@article{spaldin2019advances,
  title={Advances in magnetoelectric multiferroics},
  author={Spaldin, Nicola A and Ramesh, Rammamoorthy},
  journal={Nature Materials},
  volume={18},
  number={3},
  pages={203--212},
  year={2019},
  publisher={Nature Publishing Group UK London}
}

@article{ahn20202d,
  title={2D materials for spintronic devices},
  author={Ahn, Ethan C},
  journal={npj 2D Materials and Applications},
  volume={4},
  number={1},
  pages={17},
  year={2020},
  publisher={Nature Publishing Group UK London}
}

@article{long2020persistence,
  title={Persistence of magnetism in atomically thin MnPS3 crystals},
  author={Long, Gen and Henck, Hugo and Gibertini, Marco and Dumcenco, Dumitru and Wang, Zhe and Taniguchi, Takashi and Watanabe, Kenji and Giannini, Enrico and Morpurgo, Alberto F},
  journal={Nano Letters},
  volume={20},
  number={4},
  pages={2452--2459},
  year={2020},
  publisher={ACS Publications}
}

@article{kim2019antiferromagnetic,
  title={Antiferromagnetic ordering in van der Waals 2D magnetic material MnPS3 probed by Raman spectroscopy},
  author={Kim, Kangwon and Lim, Soo Yeon and Kim, Jungcheol and Lee, Jae Ung and Lee, Sungmin and Kim, Pilkwang and Park, Kisoo and Son, Suhan and Park, Cheol Hwan and Park, Je Geun and others},
  journal={2D Materials},
  volume={6},
  number={4},
  pages={041001},
  year={2019},
  publisher={IOP Publishing}
}

@article{taylor2011antiferromagnetic,
  title={Antiferromagnetic spin fluctuations in LiFeAs observed by neutron scattering},
  author={Taylor, AE and Pitcher, MJ and Ewings, RA and Perring, TG and Clarke, SJ and Boothroyd, AT},
  journal={Physical Review B},
  volume={83},
  number={22},
  pages={220514},
  year={2011},
  publisher={APS}
}

@article{muller2013grazing,
  title={Grazing incidence small-angle neutron scattering: challenges and possibilities},
  author={M{\"u}ller-Buschbaum, Peter},
  journal={Polymer Journal},
  volume={45},
  number={1},
  pages={34--42},
  year={2013},
  publisher={Nature Publishing Group}
}

@article{reshak2017spin,
  title={Spin-polarized second harmonic generation from the antiferromagnetic CaCoSO single crystal},
  author={Reshak, AH},
  journal={Scientific Reports},
  volume={7},
  number={1},
  pages={46415},
  year={2017},
  publisher={Nature Publishing Group UK London}
}

@article{du2023electrical,
  title={Electrical manipulation and detection of antiferromagnetism in magnetic tunnel junctions},
  author={Du, Ao and Zhu, Daoqian and Cao, Kaihua and Zhang, Zhizhong and Guo, Zongxia and Shi, Kewen and Xiong, Danrong and Xiao, Rui and Cai, Wenlong and Yin, Jialiang and others},
  journal={Nature Electronics},
  volume={6},
  number={6},
  pages={425--433},
  year={2023},
  publisher={Nature Publishing Group UK London}
}

@article{jungwirth2016antiferromagnetic,
  title={Antiferromagnetic spintronics},
  author={Jungwirth, Tomas and Marti, X and Wadley, P and Wunderlich, J},
  journal={Nature Nanotechnology},
  volume={11},
  number={3},
  pages={231--241},
  year={2016},
  publisher={Nature Publishing Group}
}

@article{malard2009group,
  title={Group-theory analysis of electrons and phonons in N-layer graphene systems},
  author={Malard, Leandro M and Guimar{\~a}es, Marcos HD and Mafra, Daniela L and Mazzoni, Mario SC and Jorio, Ado},
  journal={Physical Review B},
  volume={79},
  number={12},
  pages={125426},
  year={2009},
  publisher={APS}
}

@article{bhowal2022hidden,
  title={Hidden k-space magnetoelectric multipoles in nonmagnetic ferroelectrics},
  author={Bhowal, Sayantika and Collins, Stephen P and Spaldin, Nicola A},
  journal={Physical Review Letters},
  volume={128},
  number={11},
  pages={116402},
  year={2022},
  publisher={APS}
}

@article{vsmejkal2022emerging,
  title={Emerging research landscape of altermagnetism},
  author={{\v{S}}mejkal, Libor and Sinova, Jairo and Jungwirth, Tomas},
  journal={Physical Review X},
  volume={12},
  number={4},
  pages={040501},
  year={2022},
  publisher={APS}
}

@article{bhowal2024ferroically,
  title={Ferroically ordered magnetic octupoles in d-wave altermagnets},
  author={Bhowal, Sayantika and Spaldin, Nicola A},
  journal={Physical Review X},
  volume={14},
  number={1},
  pages={011019},
  year={2024},
  publisher={APS}
}

@article{spaldin2008toroidal,
  title={The toroidal moment in condensed-matter physics and its relation to the magnetoelectric effect},
  author={Spaldin, Nicola A and Fiebig, Manfred and Mostovoy, Maxim},
  journal={Journal of Physics: Condensed Matter},
  volume={20},
  number={43},
  pages={434203},
  year={2008}
}

@article{thole2018magnetoelectric,
  title={Magnetoelectric multipoles in metals},
  author={Th{\"o}le, Florian and Spaldin, Nicola A},
  journal={Philosophical Transactions of the Royal Society A: Mathematical, Physical and Engineering Sciences},
  volume={376},
  number={2134},
  year={2018},
  publisher={The Royal Society}
}

@article{jeevan2011muon,
  title={Muon spin relaxation and neutron diffraction investigations of quadrupolar and magnetically ordered states of YbRu 2 Ge 2},
  author={Jeevan, HS and Adroja, DT and Hillier, AD and Hossain, Z and Ritter, C and Geibel, C},
  journal={Physical Review B},
  volume={84},
  number={18},
  pages={184405},
  year={2011},
  publisher={APS}
}

@article{dugdale2014probing,
  title={Probing the Fermi surface by positron annihilation and Compton scattering},
  author={Dugdale, SB},
  journal={Low Temperature Physics},
  volume={40},
  number={4},
  pages={328--338},
  year={2014},
  publisher={AIP Publishing}
}

@incollection{mills1995positron,
  title={Positron and positronium emission spectroscopies},
  author={Mills Jr, AP},
  booktitle={Positron Spectroscopy of Solids},
  pages={209--258},
  year={1995},
  publisher={IOS Press}
}

@article{lin2020dirac,
  title={Dirac fermions in antiferromagnetic FeSn kagome lattices with combined space inversion and time-reversal symmetry},
  author={Lin, Zhiyong and Wang, Chongze and Wang, Pengdong and Yi, Seho and Li, Lin and Zhang, Qiang and Wang, Yifan and Wang, Zhongyi and Huang, Hao and Sun, Yan and others},
  journal={Physical Review B},
  volume={102},
  number={15},
  pages={155103},
  year={2020},
  publisher={APS}
}

@article{tao2024layer,
  title={Layer Hall detection of the N{\'e}el vector in centrosymmetric magnetoelectric antiferromagnets},
  author={Tao, LL and Zhang, Qin and Li, Huinan and Zhao, Hong Jian and Wang, Xianjie and Song, Bo and Tsymbal, Evgeny Y and Bellaiche, Laurent},
  journal={Physical Review Letters},
  volume={133},
  number={9},
  pages={096803},
  year={2024},
  publisher={APS}
}

@article{vsmejkal2022beyond,
  title={Beyond conventional ferromagnetism and antiferromagnetism: A phase with nonrelativistic spin and crystal rotation symmetry},
  author={{\v{S}}mejkal, Libor and Sinova, Jairo and Jungwirth, Tomas},
  journal={Physical Review X},
  volume={12},
  number={3},
  pages={031042},
  year={2022},
  publisher={APS}
}

@article{cooper1985compton,
  title={Compton scattering and electron momentum determination},
  author={Cooper, Malcom J},
  journal={Reports on Progress in Physics},
  volume={48},
  number={4},
  pages={415--481},
  year={1985}
}

@article{bandyopadhyay2025designing,
  title={Designing nonrelativistic spin splitting in oxide perovskites},
  author={Bandyopadhyay, Subhadeep and Picozzi, Silvia and Bhowal, Sayantika},
  journal={Physical Review B},
  volume={112},
  number={6},
  pages={064405},
  year={2025},
  publisher={APS}
}

@article{xue2025reversing,
  title={Reversing N{\'e}el Vector in Parity-Time Antiferromagnets by Nonreciprocal Light Scattering},
  author={Xue, Qianqian and Zhou, Jian},
  journal={Nano Letters},
  volume={25},
  number={22},
  pages={9054--9060},
  year={2025},
  publisher={ACS Publications}
}

@article{shao2020nonlinear,
  title={Nonlinear anomalous Hall effect for N{\'e}el vector detection},
  author={Shao, Ding Fu and Zhang, Shu Hui and Gurung, Gautam and Yang, Wen and Tsymbal, Evgeny Y},
  journal={Physical Review Letters},
  volume={124},
  number={6},
  pages={067203},
  year={2020},
  publisher={APS}
}
\end{document}